\documentclass[letterpaper,twocolumn,showpacs,pra,aps]{revtex4-1}

\usepackage{amsmath}  
\usepackage{amsfonts} 
\usepackage{graphicx} 
\usepackage{amssymb}

\begin{document}

\title{Enhancement of superconductivity outside an Abrikosov vortex core in a tightly bound Cooper pair superconductor}

\author{Eugene B. Kolomeisky{*}, Mia Kyler and Ishaan U. Patel}

\affiliation{Department of Physics, University of Virginia, P. O. Box 400714, Charlottesville, Virginia 22904-4714, USA\\ *email: \href{mailto:ek6n@virginia.edu}{ek6n@virginia.edu}}

\date{\today}

\begin{abstract}
Abrikosov vortices play a central role in the disruption of superconductivity in type-II superconductors. It is commonly accepted that as one moves away from the vortex's axis of an $s$-wave superconductor, the density of superconductive electrons gradually increases from zero to its bulk value. However, we demonstrate that this behavior is qualitatively altered in the zero-temperature limit provided that the Cooper pairs comprising the superconductive liquid are sufficiently tightly bound. Specifically, outside the vortex core, the density of superconductive electrons reaches a maximum surpassing its bulk value.  This phenomenon has electrostatic origins: since normal electrons are absent and there exists a charged ionic background, the spatial variation of the charge density of superconductive electrons violates local neutrality, leading to the generation of an electric field. This electric field shrinks the vortex core and turns the density profile into that with a maximum, ensuring global neutrality.  The effect is most pronounced in the limit of strong electrostatic screening, where the field configurations describing the vortex attain a universal form, with the electric field screened over a length scale determined by the London penetration depth.

\end{abstract}


\maketitle

\section{Introduction}

The understanding of superconductivity, one of the macroscopic manifestations of quantum mechanics, is shaped by the phenomenological theories of F. and H. London \cite{London,Londonv1}, the Ginzburg-Landau theory, and the microscopic theory of Bardeen, Cooper, Schrieffer, and Bogoliubov \cite{BCS,Bogo1,LL9}.  

One of the defining characteristics of the superconductive state is the existence of topological defects known as Abrikosov flux lines (or vortices). Cores of these vortices (in conventional $s$-wave superconductors neglecting spin-orbit interaction \cite{Volovik}) may be viewed as line inclusions of the normal phase traversing a sample. Consequently, they play a pivotal role in the gradual degradation of superconductivity due to an external magnetic field in type-II superconductors, as elucidated in the pioneering work of Abrikosov \cite{Abrikosov, LL9}. 

Individual flux lines are classical topological soliton solutions that can be derived from both the Ginzburg-Landau theory (which is limited to the vicinity of the transition to the superconductive state) and the London theory (which does not have such a limitation). These flux lines possess the following properties \cite{Abrikosov,LL9}:  

As one moves away from the vortex axis, the superconductive density gradually increases from zero to its bulk value. Most of this change occurs within the vortex core. Additionally, there is a monotonically decreasing magnetic field profile.  

The objective of this study is to demonstrate that this picture needs a qualitative alteration in the zero-temperature limit, which can be understood as follows:  
 
Two types of electrons - superconductive and normal - are known to coexist in a metal superconductor. Superconductive electrons formed by electrons bound in Cooper pairs participate in the equilibrium superconductivity current.  On the other hand, the current of normal electrons is non-equilibrium and dissipative. As the temperature is lowered below the transition point into the superconductive state, the density of superconductive electrons begins to increase while the density of normal electrons starts to decrease. This behavior persists as the temperature is further reduced, so that at zero temperature all the electrons are superconductive. 

Since normal electrons are abundant near the transition point into the superconductive state, a non-uniform density profile of superconductive electrons (like in the Abrikosov vortex) does not violate local neutrality. This is the region where the Ginzburg-Landau theory applies, and electrostatic effects are absent.  The density profile of superconductive electrons in a vortex is a monotonically increasing function of distance from its axis \cite{Abrikosov,LL9}.  

On the other hand, in the zero-temperature limit, the presence of a charged background in the form of an ionic lattice, the absence of normal electrons, and a depletion of superconductive electrons in the vortex core region lead to a violation of local neutrality. Consequently, an outward-directed radial electric field is generated, pulling superconductive electrons closer to the vortex axis, thus screening the lattice charge. Since the lowest energy state is that of global neutrality, the density profile can no longer be a monotonically increasing function of the distance from the vortex axis. Instead, it becomes a non-monotonic function that is expected to have a single maximum.  The bulk value of the superconductive density is now approached from above as the distance from the vortex axis increases. Moreover, there will be associated changes in the profile of the magnetic field.  Electrostatic effects are relevant and the above reasoning holds if Cooper pairs are sufficiently tightly bound so that gradients of electric fields are not strong enough to break them into normal electrons.  

Electrostatic effects can be included via a phenomenological theory due to Feynman \cite{F1,F2}. A self-contained description of this theory is given below followed by an analysis of Abrikosov-type vortex solutions. 

\section{Feynman's phenomenological theory}

The basis of this approach lies in an insight \cite{F1,F2} that Schr\"odinger's interpretation of the square of the wave function's magnitude, $|\Psi(\mathbf{r},t)|^{2}$, as proportional to the charge density at location $\mathbf{r}$ and time $t$, holds true in superconductors. This is because electron Cooper pairs, which are Bose particles, can occupy the same state in large numbers, similar to the situation in the Ginzburg-Landau theory \cite{LL9}. Consequently, the superconductive state should be described by a Schr\"odinger equation, where the wave function has a classical interpretation.

Feynman's treatment of superconductivity is an example of a classical non-relativistic field theory that begins with the action of the form
\begin{equation}
\label{variational}
S=\int \mathcal{L}\left (q,\frac{\partial q}{\partial t},\nabla q,\mathbf{r},t\right )dtd^{3}x
\end{equation}
where $\mathcal{L}$ is the Lagrangian density (hereafter called the Lagrangian) and $q$ is a shorthand for a full set of dynamical field quantities describing the state of the system.  They include the real and imaginary parts of the wave function $\Psi$ (or equivalently $\Psi$ and $\Psi^{*}$), the scalar potential and components of the vector potential.  

The principle of least action $\delta S=0$, i.e. the requirement that variation of $\mathcal{L}$ with respect to $q$ is zero, $\delta\mathcal{L}/\delta q=0$, gives the Euler-Lagrange (field) equations of the form \cite{F2,Schiff}:
\begin{equation}
\label{Euler_Lagrange}
\frac{\partial}{\partial t}\frac{\partial \mathcal{L}}{\partial (\partial_{t}q)}+\sum_{j=1}^{3}\frac{\partial}{\partial x_{j}}\frac{\partial \mathcal{L}}{\partial (\partial_{j}q)}-\frac{\partial \mathcal{L}}{\partial q}=0
\end{equation}
where $x_{1}\equiv x$, $x_{2}\equiv y$, and $x_{3}\equiv z$.  In what follows we will also need an expression for the energy of the field configuration encoded in $q$ \cite{Schiff}
\begin{equation}
\label{general_energy}
\mathcal{E}=\int\left [\sum_{q}\frac{\partial q}{\partial t}\frac{\partial \mathcal{L}}{\partial (\partial_{t}q)}-\mathcal{L}\right ]d^{3}x
\end{equation}
where the summation is performed over all the fields belonging to the set $q$.

It is instructive to start with the case of a superfluid of neutral particles of mass $m$.  The particles may be "elementary" bosons (such as $^{4}$\textrm{He} atoms) or "composite" bosons made of pairs of bound neutral spin-1/2 fermions.  This system is described by the Lagrangian   
\begin{equation}
\label{neutral_Lagrangian}
\mathcal{L}=\frac{i\hbar}{2}\left (\Psi^{*}\frac{\partial\Psi}{\partial t}-\Psi\frac{\partial \Psi^{*}}{\partial t}\right )
-\frac{\hbar^{2}}{2m}|\nabla\Psi|^{2}-V\left (|\Psi|^{2}\right )
\end{equation}
where $n(\mathbf{r},t) =|\Psi(\mathbf{r},t)|^{2}$ has the classical interpretation of the superfluid number density and the function $V\left (|\Psi|^{2}\right )$ represents interparticle interactions.  If $V=0$ then Eq.(\ref{neutral_Lagrangian}) reduces to the Lagrangian of the linear Schr\"odinger field theory \cite{Schiff}.  

In what follows we find it useful to employ the Madelung representation of the wave function \cite{Madelung}
\begin{equation}
\label{Madelung}
\Psi(\mathbf{r},t)=\sqrt{n(\mathbf{r},t)}e^{i\theta(\mathbf{r},t)}
\end{equation}
where $\theta$ is the phase of the wave function.  The hallmark of the superfluid (and superconductive) state is an interaction function $V(n)$ that has a minimum at a finite value of $n=n_{0}$.  For $n$ close to $n_{0}$ the interaction $V(n)$ can be written as 
\begin{equation}
\label{minimum}
V(n)=\frac{1}{2}\frac{mu^{2}}{n_{0}}(n-n_{0})^{2},~~~u^{2}=\frac{n_{0}}{m}V''(n_{0})>0
\end{equation}  
where the parameter $u(n_{0})$ is the velocity of sound.  We can now introduce the coherence length 
\begin{equation}
\label{coherence_length}
\xi=\frac{\hbar}{\sqrt{2}mu}
\end{equation} 
that has a meaning of the healing length of a small density disturbance away from the equilibrium value $n=n_{0}$.

We now turn attention to the case of a superconductor of Cooper pairs of net charge $e$ and effective mass $m$ and use the treatment of the neutral superfluid as a template on which a theory of superconductivity can be built.  The function $n(\mathbf{r},t)$ in Eq.(\ref{Madelung}) is then the superconductive density while the parameter $n_{0}$ in Eq.(\ref{minimum}) is the equilibrium superconductive density.

(i) In order to couple the Schr\"odinger field $\Psi$ representing charged particles to the electromagnetic field the derivatives of the wave function $\Psi$ have to be modified in a manner guaranteeing gauge invariance \cite{Schiff}:
\begin{equation}
\label{minimal_coupling}
\frac{\partial\Psi}{\partial t}\rightarrow \left (\frac{\partial}{\partial t}+\frac{ie}{\hbar}\phi\right )\Psi,~~~\nabla\Psi\rightarrow\left (\nabla-\frac{ie}{\hbar c}\mathbf{A}\right )\Psi
\end{equation} 
where $\phi$ and $\mathbf{A}$ are the scalar and vector potentials defining the electric $\mathbf{E}$ and magnetic $\mathbf{H}$ fields 
\begin{equation}
\label{EM_fieds}
\mathbf{E}=-\nabla\phi-\frac{1}{c}\frac{\partial\mathbf{A}}{\partial t},~~~\mathbf{H}=\nabla\times\mathbf{A}.
\end{equation}

(ii) In the zero temperature limit there are only superconductive electrons present and the equilibrium superconductive density $n_{0}$ is fixed by electrostatics - it is the density of the ionic lattice guaranteeing local charge neutrality.  To account for the presence of this static charged background, the Lagrangian must have an additional $en_{0}\phi$ term.  

(iii) The Lagrangian of the electromagnetic field is $(\mathbf{E}^{2}-\mathbf{H}^{2})/8\pi$ \cite{Schiff}.

Accounting for all these provisions, the Lagrangian describing the superconductor acquires the form
\begin{eqnarray}
\label{charged_Lagrangian}
\mathcal{L}&=&\frac{i\hbar}{2}\left (\Psi^{*}\frac{\partial\Psi}{\partial t}-\Psi\frac{\partial\Psi^{*}}{\partial t}\right )-e\phi\left (|\Psi|^{2}-n_{0}\right )\nonumber\\
&-&\frac{\hbar^{2}}{2m} \left | \left (\nabla-\frac{ie}{\hbar c}\mathbf{A}\right )\Psi\right |^{2}
-V\left (|\Psi|^{2}\right )\nonumber\\
&+&\frac{1}{8\pi}\left (\frac{1}{c}\frac{\partial \mathbf{A}}{\partial t}+\nabla\phi\right )^{2}-\frac{1}{8\pi}(\nabla\times\mathbf{A})^{2}
\end{eqnarray}
that is invariant with respect to a local gauge transformation:
\begin{eqnarray}
\label{local_gauge}
\phi&\rightarrow&\phi'=\phi-\frac{1}{c}\frac{\partial \chi(\mathbf{r},t)}{\partial t},~\mathbf{A}\rightarrow\mathbf{A}'=\mathbf{A}+\nabla\chi(\mathbf{r},t)\nonumber\\\Psi&\rightarrow&\Psi'=\Psi e^{ie\chi(\mathbf{r},t)/\hbar c}
\end{eqnarray}
where $\chi(\mathbf{r},t)$ is an arbitrary real function.  For $V=0$ and $n_{0}=0$ the Lagrangian was given in Ref.\cite{Schiff}.  The presence of the interaction term $V$ of the form (\ref{minimum}) (for $n$ close to $n_{0}$) and the meaning of the parameter $n_{0}$ were discussed by Feynman \cite{F1,F2}.  A microscopic derivation of the Lagrangian of a superconductor (\ref{charged_Lagrangian}) is also available \cite{micro}.

Requiring that  $\delta \mathcal{L}/\delta\Psi^{*}=0$, applied to Eq.(\ref{charged_Lagrangian}), leads, according to Eq.(\ref{Euler_Lagrange}), to the nonlinear Schr\"odinger equation 
\begin{equation}
\label{GL_equation}
i\hbar\left (\frac{\partial}{\partial t}+\frac{ie}{\hbar}\phi\right )\Psi=\bigg\{-\frac{\hbar^{2}}{2m}\left (\nabla-\frac{ie}{\hbar c}\mathbf{A}\right )^{2}+V'\left (|\Psi|^{2}\right )\bigg\}\Psi.
\end{equation}
In the Madelung representation (\ref{Madelung}) it turns into two equations.  One of them is the equation of continuity
\begin{equation}
\label{continuity}
\frac{\partial n}{\partial t} +\nabla \cdot(n\textbf{v})=0,~~\mathbf{v}=\frac{\hbar}{m}\left (\nabla\theta-\frac{e}{\hbar c}\mathbf{A}\right )
\end{equation}
where $\mathbf{v}$ is the flow velocity of the superconductive liquid.   It satisfies the relationship
\begin{equation}
\label{rotational}
\nabla\times\mathbf{v}=-\frac{e}{mc}\nabla\times\mathbf{A}=-\frac{e}{mc}\mathbf{H}
\end{equation}
conjectured by F. London \cite{Londonv1};  here it is a consequence of the gauge invariance of the underlying Lagrangian (\ref{charged_Lagrangian}). 

The second equation arising from the Madelung substitution (\ref{Madelung}) into the nonlinear Schr\"odinger equation (\ref{GL_equation}) is
\begin{equation}
\label{Bernoulli_modified}
-\hbar\left (\frac{\partial\theta}{\partial t}+\frac{e}{\hbar}\phi\right )=\frac{m\mathbf{v}^{2}}{2}+V'(n)-\frac{\hbar^{2}}{2m}\frac{\nabla^{2}\sqrt{n}}{\sqrt{n}}.
\end{equation}    

Requiring that $\delta \mathcal{L}/\delta\phi=0$, applied to Eq.(\ref{charged_Lagrangian}), leads, according to Eq.(\ref{Euler_Lagrange}), to Gauss's law of electricity
\begin{equation}
\label{electric_Gauss}
\nabla\cdot \mathbf{E}=4\pi e\left (n-n_{0}\right ).
\end{equation}
Likewise, imposing $\delta \mathcal{L}/\delta\mathbf{A}=0$ results in Amp\`ere-Maxwell's law
\begin{equation}
\label{AM}
\nabla\times \textbf{H}=\frac{1}{c}\frac{\partial \textbf{E}}{\partial t}+\frac{4\pi}{c}en\mathbf{v}.
\end{equation}
As is always the case in electrodynamics, the continuity equation (\ref{continuity}) is simultaneously a consequence of Gauss's (\ref{electric_Gauss}) and  Amp\`ere-Maxwell's (\ref{AM}) laws.

Finally, employing Eqs.(\ref{general_energy}), (\ref{Madelung}), (\ref{charged_Lagrangian}), and (\ref{electric_Gauss}) as well as the divergence theorem, the energy can be computed as
\begin{equation}
\label{superconductor_energy}
\mathcal{E}=\int\left [\frac{\hbar^{2}}{2m}(\nabla\sqrt{n})^{2}+\frac{mn\mathbf{v}^{2}}{2}+V(n)+\frac{\mathbf{E}^{2}+\mathbf{H}^{2}}{8\pi}\right ]d^{3}x.
\end{equation} 

\section{Stationary field equations}

The Abrikosov flux line is described by the stationary version of the field equations that include:

(i)  The stationary versions of the continuity equation (\ref{continuity})
\begin{equation}
\label{stationary_continuity}
\nabla \cdot(n\textbf{v})=0.
\end{equation}

(ii) The stationary version of Eq.(\ref{Bernoulli_modified}) 
\begin{equation}
\label{stationary_Bernoulli_modified}
\frac{m\mathbf{v}^{2}}{2}+e\phi+\frac{mu^{2}}{n_{0}}(n-n_{0})-\frac{\hbar^{2}}{2m}\frac{\nabla^{2}\sqrt{n}}{\sqrt{n}}=0.
\end{equation}
Hereafter without significant loss of generality we assume that the function $V(n)$ characterizing interactions has the parabolic form (\ref{minimum}) for \textit{all} $n$.  This choice also allows us to make direct comparison with conclusions of the Abrikosov theory \cite{Abrikosov,LL9}. 

(iii) The Poisson equation
\begin{equation}
\label{Poisson}
\nabla^{2}\phi=-4\pi e(n-n_{0}).
\end{equation}  

(iv) The Amp\`ere law limit of Eq.(\ref{AM})
\begin{equation}
\label{stationary_AM}
\nabla\times \textbf{H}=\frac{4\pi}{c}en\mathbf{v}. 
\end{equation}

\section{Abrikosov flux line}

We will be seeking $z$-independent solutions to these equations that minimize the energy (\ref{superconductor_energy}) per unit length (also called the line energy):
\begin{eqnarray}
\label{superconductor_energy_per_unit_length}
\varepsilon=\frac{\mathcal{E}}{L_{z}}&=&\int\Big\{\frac{\hbar^{2}}{2m}(\nabla\sqrt{n})^{2}+\frac{mu^{2}}{2n_{0}}(n-n_{0})^{2}\nonumber\\
&+&\frac{mn\mathbf{v}^{2}}{2}+\frac{\mathbf{E}^{2}+\mathbf{H}^{2}}{8\pi}\Big\}dxdy
\end{eqnarray}
where $L_{z}$ is macroscopic system size along the $z$ direction.  Hereafter the cylindrical system of coordinates $\rho, \varphi,z$ with the vortex axis coinciding with the $z$-axis will be employed. 

Since the integrand is a sum of positively-defined contributions, the necessary (but not sufficient) condition for the minimum of (\ref{superconductor_energy_per_unit_length}) is that every one of its integrand entries vanishes as $\rho\rightarrow\infty$.  

When applied to the first two terms involving $n(\rho,\varphi)$, this criterion implies that $n(\rho,\varphi)\rightarrow n_{0}$ as $\rho\rightarrow\infty$, and the wave function (\ref{Madelung}) has the limiting behavior of the form $\Psi(\rho\rightarrow\infty,\varphi)=\sqrt{n_{0}}\exp[ i\theta(\varphi)]$.  The constraint that it must be single-valued then dictates that      
\begin{equation}
\label{single-valuedness}
\theta(\varphi+2\pi)-\theta(\varphi)=2\pi l,~~~l=0,\pm 1,\pm 2,...
\end{equation}

The requirement that $mn\mathbf{v}^{2}/2$, the kinetic energy density term in (\ref{superconductor_energy_per_unit_length}), vanishes as $\rho\rightarrow\infty$ combined with the field equation (\ref{stationary_Bernoulli_modified}) implies that the scalar potential $\phi(\rho\rightarrow\infty)$ also vanishes.  Then $\phi(\rho)$ is a direct measure of electric effects;  it is the electrostatic potential difference at a distance $\rho$ away from the vortex axis and infinity.

Employing the definition of the flow velocity (\ref{continuity}) one also infers that
\begin{equation}
\label{A_large_rho}
\mathbf{A}=\frac{\hbar c}{e}\nabla \theta=\frac{\hbar c}{e}\frac{1}{\rho}\frac{d\theta}{d \varphi}\mathbf{e}_{\varphi}, ~\text{as}~\rho\rightarrow\infty
\end{equation}
(hereafter $\mathbf{e}_{\rho,\varphi,z}$ are unit vectors of the cylindrical system of coordinates).  As a result the magnetic field $\mathbf{H}=\nabla\times\mathbf{A}$ vanishes as $\rho\rightarrow\infty$.

Eqs.(\ref{single-valuedness}) and (\ref{A_large_rho}) also imply that the magnetic flux over the entire $xy$ plane is quantized in units of $\Phi_{0}=2\pi\hbar c/|e|$, the flux quantum \cite{LL9}:   
\begin{equation}
\label{flux_quantization}
\int \mathbf{H}\cdot d\mathbf{s}=\oint \mathbf{A}\cdot d\mathbf{l}
=\frac{\hbar c}{e}\oint \nabla\theta\cdot d\mathbf{l}=\frac{2\pi\hbar c}{e}l
\end{equation}
where $d\mathbf{s}=\mathbf{e}_{z}dxdy$ and the Stokes theorem was employed in the second step.  

In the same $\rho\rightarrow\infty$ limit the continuity equation (\ref{stationary_continuity}) becomes $\nabla(\nabla\theta-e\mathbf{A}/\hbar c)=0$.  It further simplifies in the Coulomb gauge $\nabla\cdot\mathbf{A}=0$, becoming the Laplace equation $\nabla^{2}\theta=0$.  Its relevant solution consistent with the condition of single-valuedness (\ref{single-valuedness}) is $\theta=l\varphi$.

These considerations motivate the following ansatz for the solution to the stationary field equations for all $\rho$:
\begin{eqnarray}
\label{conjecture}
\Psi&=&\sqrt{n(\rho)}e^{i\theta(\varphi)}\equiv\sqrt{n_{0}}R(\rho)e^{il\varphi},\nonumber\\
\mathbf{A}&=&\frac{\hbar cl}{e}\frac{1-F(\rho)}{\rho}\mathbf{e}_{\varphi}, \phi=\frac{mu^{2}}{e}w(\rho).
\end{eqnarray}
Here $R^{2}(\rho)$ is the dimensionless density satisfying the boundary conditions $R(\infty)=1$ and $R(0)=0$ where the latter is a consequence of the single-valuedness of the macroscopic wave function in Eq.(\ref{conjecture}) at $\rho=0$.

The dimensionless function $F(\rho)$ entering the expression for the vector potential in Eq.(\ref{conjecture}) satisfies the boundary conditions $F(\rho\rightarrow\infty)\rightarrow 0$ and $F(\rho\rightarrow 0)\rightarrow 1$ (to prevent singularity of the vector potential at $\rho=0$).  It also determines the behavior of the flow velocity (\ref{continuity}):  
\begin{equation}
\label{vortex_velocity}
\mathbf{v}=\frac{\hbar l}{m}\frac{F(\rho)}{\rho}\mathbf{e}_{\varphi}.
\end{equation}  
Since both the density and flow velocity depend only on $\rho$ while the flow velocity is tangential, $\mathbf{v}\propto\mathbf{e}_{\varphi}$, the continuity equation (\ref{stationary_continuity}) is automatically satisfied.

The function $F(\rho)$ additionally determines the behavior of the magnetic field
\begin{equation}
\label{magnetic_field}
\mathbf{H}=\nabla\times\mathbf{A}=H(\rho)\mathbf{e}_{z}, ~~H(\rho)=-\frac{\hbar cl}{e}\frac{1}{\rho}\frac{dF}{d\rho}
\end{equation}
directed along the $z$-axis.  

The function $w(\rho)$ (hereafter referred to as the potential) in Eq. (\ref{conjecture}) is the dimensionless electrostatic potential energy whose zero is set at infinity.  The electric field is radial.

We can now introduce two additional length scales, the London penetration depth $\lambda$ and the Debye screening length $\delta$,
\begin{equation}
\label{London_Debye}
\lambda^{2}=\frac{mc^{2}}{4\pi n_{0}e^{2}},~~~\delta^{2}=\frac{mu^{2}}{4\pi n_{0}e^{2}},
\end{equation}     
that quantify the ranges of magnetic and electrostatic screening, respectively.

\subsection{Field configuration}

Substituting Eqs.(\ref{conjecture}), (\ref{vortex_velocity}) and (\ref{magnetic_field}) into the field equations (\ref{stationary_Bernoulli_modified}), (\ref{Poisson}), (\ref{stationary_AM}), and measuring length in units of the coherence length $\xi$ (\ref{coherence_length}) we arrive at a coupled system of three nonlinear second-order differential equations for the three unknown functions $R(\rho)$, $w(\rho)$, and $F(\rho)$:
\begin{eqnarray}
\label{Bernoulli_dimensionless}
\frac{l^{2}F^{2}}{\rho^{2}}-\frac{1}{\rho R}\frac{d}{d\rho}\left (\rho\frac{dR}{d\rho}\right )&+&R^{2}-1+w=0,\nonumber\\
R(0)&=&0,~~~R(\infty)=1;
\end{eqnarray}
\begin{equation}
\label{Poisson_dimensionless}
\frac{\gamma^{2}}{\rho}\frac{d}{d\rho}\left (\rho\frac{dw}{d\rho}\right )=1-R^{2},~~w(\infty)=0,~ \gamma=\frac{\delta}{\xi};
\end{equation}
\begin{equation}
\label{Ampere_dimensionless}
\kappa^{2}\rho\frac{d}{d\rho}\left (\frac{1}{\rho}\frac{dF}{d\rho}\right )=R^{2}F,~F(0)=1,~F(\infty)=0,~\kappa=\frac{\lambda}{\xi}
\end{equation} 
where $\kappa$, the dimensionless London penetration depth (\ref{London_Debye}), is the zero-temperature counterpart of the Ginzburg-Landau parameter \cite{LL9}.  The parameter $\gamma$ is the dimensionless Debye screening length (\ref{London_Debye}).  

Since in a metal superconductor the equilibrium superconductive density $n_{0}$ has the order of magnitude of the electron density in a normal metal while the effective mass $m$ has the order of magnitude of the electron mass, the parameters $\kappa$ and $\gamma$ can be estimated as
\begin{equation}
\label{ kappa_gamma}
\kappa\simeq\frac{u}{\alpha^{2}c}, ~~~\gamma\simeq\frac{u^{2}}{\alpha^{2}c^{2}}
\end{equation}      
where $\alpha=(e/2)^{2}/\hbar c\approx 1/137$ is the fine structure constant.  Assuming $c/u\approx 10^2$ (i.e. the speed of sound $u$ has the order of magnitude of the Fermi velocity in a normal metal), we discover that $\kappa\simeq 10^{2}$ while $\gamma\simeq 1$. Restated in terms of the original length scales of the problem, this implies that the London penetration depth $\lambda$ is significantly larger than the coherence length, $\lambda\gg \xi$.  In the classification of Ginzburg and Landau and Abrikosov \cite{LL9,Abrikosov} this is a strongly type-II superconductor.  

General reasoning alone is not sufficient to establish a definite relationship between the coherence length $\xi$ and the Debye screening length $\delta$. Depending on specific material parameters, both $\xi < \delta$ and $\xi >\delta$ scenarios are possible. For instance, in a metal superconductor, the screening length $\delta$ is anticipated to be of the atomic scale. However, the coherence length $\xi$ can range from macroscopic to atomic scales. Consequently, the $\gamma\lesssim 1$ (strong screening) case appears more plausible than $\gamma \gtrsim 1$ (weak screening) one.            

Multiplying both sides of the Poisson equation (\ref{Poisson_dimensionless}) by $\rho$, assuming the electric field, falls off with distance faster than $1/\rho$, and integrating from zero to infinity, we obtain
\begin{equation}
\label{global neutrality}
\int_{0}^{\infty}(1-R^{2})\rho d\rho=0
\end{equation}
which is the condition of global neutrality. 

All previously studied cases of vortices are encompassed by Eqs.(\ref{Bernoulli_dimensionless}), (\ref{Poisson_dimensionless}) and (\ref{Ampere_dimensionless}). Specifically, setting $\gamma=\infty$ we find that $w=0$, i.e. electrostatic screening is absent on all scales.  Then Eqs.(\ref{Bernoulli_dimensionless}) and (\ref{Ampere_dimensionless}) reduce to those due to Abrikosov \cite{Abrikosov}.  The structure of the Abrikosov vortex line encoded in functions $R(\rho)$ and $F(\rho)$ is solely determined by the parameter $\kappa$.  In the original analysis \cite{Abrikosov} Abrikosov used the function $F/\rho$ instead of $F$ and measured length in units of the London penetration depth $\lambda$ (\ref{London_Debye}).   

In order to take non-relativistic or equivalently electrostatic limit one has to revisit the starting point of the analysis as the notion of the vector potential is no longer needed.  Indeed, we will be seeking solutions to the field equations (\ref{stationary_continuity}), (\ref{stationary_Bernoulli_modified}) and (\ref{Poisson}) minimizing the energy per unit length (\ref{superconductor_energy_per_unit_length}) omitting the magnetic energy density $\mathbf{H}^{2}/8\pi$ term.  While this has no effect on the condition of single-valuedness of the wave function (\ref{single-valuedness}) and the boundary condition $\phi(\rho\rightarrow\infty)=0$, Eq.(\ref{A_large_rho}) is replaced by that for the flow velocity (\ref{continuity})
\begin{equation}
\label{v_large_rho}
\mathbf{v}=\frac{\hbar}{m}\nabla \theta\rightarrow\frac{\hbar}{m\rho}\frac{d\theta}{d \varphi}\mathbf{e}_{\varphi}, ~\text{as}~\rho\rightarrow\infty.
\end{equation}
Its immediate consequence is a quantization of the velocity circulation over a path encircling the $xy$ plane:
\begin{equation}
\label{circulation_quantization}
\oint \mathbf{v}\cdot d\mathbf{l}=\frac{\hbar}{m}\oint \nabla\theta\cdot d\mathbf{l}=\frac{2\pi\hbar }{m}l.
\end{equation}
In the same limit the continuity equation (\ref{stationary_continuity}) becomes the Laplace equation $\nabla^{2}\theta=0$ whose solution consistent with the condition of single-valuedness (\ref{single-valuedness}) is $\theta=l\varphi$.

These considerations imply that the ansatz for the solution to the stationary field equations in the non-relativistic limit still has the form (\ref{conjecture}) with the exception of the expression for the vector potential that will be replaced by the $F=1$ case of Eq.(\ref{vortex_velocity}).   Then the velocity circulation (\ref{circulation_quantization}) is quantized for an arbitrary contour encircling the $\rho=0$ singularity \cite{LL9}.  The vortex line is now described by the $F=1$ version of Eq.(\ref{Bernoulli_dimensionless}) along with the Poisson equation (\ref{Poisson_dimensionless}).  Its structure encoded in functions $R(\rho)$ and $w(\rho)$ is solely determined by the dimensionless Debye screening length $\gamma$ (\ref{coherence_length}).    

One can arrive at the conclusion $F=1$ more formally by taking the $\kappa\rightarrow\infty$ limit in the Amp\`ere law (\ref{Ampere_dimensionless}).  One of its two independent solutions $F=const$ corresponds to potential rotation \cite{LL9} and the first boundary condition in Eq.(\ref{Ampere_dimensionless}) dictates that $F=1$.  The second boundary condition $F(\infty)=0$ cannot be satisfied due to the lack of magnetic screening ($\kappa=\infty$).   

The $F=1$ case of Eq.(\ref{Bernoulli_dimensionless}) and Poisson's equation (\ref{Poisson_dimensionless}) describe a vortex line in the so-called non-linear Schr\"odinger-Poisson system \cite{Shuk1,Shuk2}.   Physical reasoning given at the beginning of this work equally applies here leading to the conclusion that the density profile $R^{2}(\rho)$ is a non-monotonic function with a maximum;  for $\rho$ large it approaches the bulk value $R^{2}=1$ from above.  While this result was already found numerically for a particular value of $\gamma$ \cite{Shuk1,Shuk2}, its physical explanation was lacking.

Finally, if we now exclude electrostatic effects ($\gamma=\infty$) setting $w=0$ in the $F=1$ version of Eq.(\ref{Bernoulli_dimensionless}), the latter turns into the parameter-free Ginzburg-Pitaevskii equation describing the structure of the Onsager-Feynman vortex in neutral superfluids \cite{LL9}.        

With all the effects included, the structure of the vortex line is determined by the interplay of the two independent parameters $\gamma$ and $\kappa$.  Since $\kappa\gg 1$, for $\rho\ll \kappa$ the solution to the general problem reduces to its electrostatic limit.

\subsubsection{Small distance behavior}

Thanks to the boundary conditions $F(0)=1$ and $R(0)=0$ the $\rho\rightarrow 0$ solution to Eq.(\ref{Bernoulli_dimensionless}) is given by 
\begin{equation}
\label{small_distance_density}
R(\rho)\simeq \rho^{|l|}.
\end{equation}

The small-distance behavior of the function $F(\rho)$ can be determined by combining the boundary condition $F(0)=1$ with the physical requirement that the magnetic field (\ref{magnetic_field}) at the vortex axis $H(0)$ is finite:
\begin{equation}
\label{small_distance_F}
F(\rho)=1-\frac{e\xi^{2}H(0)}{2\hbar cl}\rho^{2}.
\end{equation}
It is straightforward to verify that that such a behavior is consistent with Eqs.(\ref{Ampere_dimensionless}) and (\ref{small_distance_density}).  The weak and strong screening estimates for $H(0)$ are given below.

In the same limit integration of the Poisson equation (\ref{Poisson_dimensionless}) gives a dependence
\begin{equation}
\label{small_distance-potential}
w(\rho)=w_{0}+\frac{\rho^{2}}{4\gamma^{2}}
\end{equation}
which is an increasing function of $\rho$.  Since $w(\infty)=0$ and there are no physical reasons for the electric field to change its sign, the potential $w(\rho)$ has to be a negative monotonically increasing function of $\rho$.  The weak and strong screening estimates for the constant $w_{0}<0$ are given below.  

Even though the interaction function $V(n)$ was chosen to have the parabolic form (\ref{minimum}) for all $n$, the small-distance behavior encapsulated in Eqs.(\ref{small_distance_density}), (\ref{small_distance_F}), and (\ref{small_distance-potential}) is independent of its form.

\subsubsection{Large distance behavior:  weak-screening limit}

In the $\rho\gg 1$ limit the first two terms in Eq.(\ref{Bernoulli_dimensionless}) can be neglected if the potential $w(\rho)$ is a slowly varying function of position, i.e. $\gamma\gg 1$.  Physically this corresponds to weak screening.  In this limit we obtain an explicit relationship between the dimensionless density profile $R^{2}(\rho)$ and the potential $w(\rho)$:
\begin{equation}
\label{TF}
R^{2}(\rho)=1-w(\rho).
\end{equation} 
This approximation is analogous to the semiclassical or Thomas-Fermi (TF) approximation in quantum mechanics \cite{Schiff}.  Since $w(\rho)<0$, one concludes that $R^{2}(\rho)>1$, i.e. for $\rho$ large the bulk value of the superconductive density $R^{2}=1$ is approached from above.  In conjunction with the small-distance behavior (\ref{small_distance_density}), this implies that the $R^{2}(\rho)$ dependence has a maximum.

Substituting the TF result (\ref{TF}) into the Poisson equation (\ref{Poisson_dimensionless}) we obtain an expression 
\begin{equation}
\label{Poisson_linear}
\frac{\gamma^{2}}{\rho}\frac{d}{d\rho}\left (\rho\frac{dw}{d\rho}\right )=w,
\end{equation}
that has a form of the Debye-H\"uckel equation arising in theories of screening in classical plasma and strong electrolytes \cite{LL5}.   Its relevant axially symmetric solution has the form
\begin{equation}
\label{Bessel}
w(\rho)=aK_{0}\left (\frac{\rho}{\gamma}\right ),~~~a\simeq -\frac{1}{\gamma^{2}}
\end{equation}
where hereafter $K_{\nu}(\zeta)$ stands for the modified Bessel function \cite{NIST} of order $\nu$.  The estimate for the pre-factor $a$ is obtained from the requirement of global neutrality:  the integral (\ref{global neutrality}) is split into two - from zero to unity and from unity to infinity.  The former was estimated by setting $R=0$ while in the latter we employed the TF result (\ref{Bessel}). 

For $\rho\gg \gamma$ the potential (\ref{Bessel}) falls off with distance exponentially, $w(\rho)\simeq \gamma^{-3/2}\rho^{-1/2}\exp(-\rho/\gamma)$ \cite{NIST},  which reflects effectiveness of screening.  Since variations of the density and potential are related by the TF relationship (\ref{TF}), the density approaches its bulk value $R^{2}=1$ exponentially with the decay length set by the Debye screening length. 

In the opposite limit $1\ll\rho\ll \gamma$ employing $K_{0}(\zeta)=- \ln \zeta$ as $\zeta\rightarrow 0$ \cite{NIST} one finds an expression
\begin{equation}
\label{Bessel_small_distance}
w(\rho)\simeq -\frac{1}{\gamma^{2}}\ln\frac{\gamma}{\rho}
\end{equation}
which has a form of the unscreened potential of a charged line: electrostatic screening is ineffective at small distances.  This result is consistent with the expectation that (in the original physical units) the region of the vortex core $\rho\lesssim \xi $ is positively charged with linear charge density of the order $-en_{0}\xi^{2}$.  

Matching the asymptotic solutions (\ref{small_distance-potential}) and (\ref{Bessel_small_distance}) at $\rho\simeq 1$ gives the estimate for the constant $w_{0}$ in Eq.(\ref{small_distance-potential}) $w_{0}\simeq -\ln\gamma/\gamma^{2}$ valid in the $\ln\gamma\gg 1$ limit.

Since electrostatic screening is ineffective at distances $\rho\ll \gamma$, maximum of the density $R^{2}$ is expected to be at $\rho\simeq \gamma$ and can be estimated from Eqs.(\ref{TF}) and (\ref{Bessel}) as $R^{2}(\gamma)-1=-w(\gamma)\simeq 1/\gamma^{2}$.  As the parameter $\gamma$ decreases, i.e. the screening gets stronger, the density maximum displaces toward the vortex axis and maximal value of the density increases.  This reasoning is general and not limited to the regime of weak electrostatic screening.   

Since $|w|\ll 1$ holds in the $\gamma\gg 1$ limit, substituting the TF result (\ref{TF}) into Amp\`ere's law (\ref{Ampere_dimensionless}) only results in a slight quantitative modification to the finding of Abrikosov's theory \cite{Abrikosov},
\begin{equation}
\label{Abrikosov}
F(\rho)= \frac{\rho}{\kappa}K_{1}\left (\frac{\rho}{\kappa}\right ).
\end{equation}   
Specifically, employing the expansion $\zeta K_{1}(\zeta)=1+(\zeta^{2}/2)\ln(\zeta/2)$ as $\zeta\rightarrow 0$ \cite{NIST}, the magnetic field (\ref{magnetic_field}) at distances $\rho\ll\kappa$ is given by the expression
\begin{equation}
\label{magnetic_field_small_rho}
H(\rho)=\frac{\hbar cl}{e\lambda^{2}}\ln\frac{\kappa}{\rho},~~~\ln\frac{\kappa}{\rho}\gg 1.
\end{equation}
When evaluated at the edge of the vortex core $\rho\simeq 1$, this expression also provides an estimate for the magnetic field at the axis of the vortex, $H(0)=(\hbar cl/e\lambda^{2})\ln\kappa$, entering Eq.(\ref{small_distance_F}).

\subsubsection{Large distance behavior:  strong-screening limit}

Analytical progress can be also made in the case of strong electrostatic screening.  Taking the limit $\gamma\rightarrow 0$ in the Poisson equation (\ref{Poisson_dimensionless}) we find $R^{2}=1$ which is the statement of local neutrality.  This transforms Amp\`ere's law (\ref{Ampere_dimensionless}) into the equation  
\begin{equation}
\label{Ampere_TF}
\kappa^{2}\rho\frac{d}{d\rho}\left (\frac{1}{\rho}\frac{dF}{d\rho}\right )=F
\end{equation}
with the solution given by Eq.(\ref{Abrikosov}).

Substituting Eq.(\ref{Abrikosov}) and the condition of local neutrality $R^{2}=1$ in Eq.(\ref{Bernoulli_dimensionless}) determines the potential 
\begin{equation}
\label{limiting_potential}
w(\rho)=-\frac{l^{2}F^{2}(\rho)}{\rho^{2}}=-\frac{l^{2}}{\kappa^{2}}K_{1}^{2}\left (\frac{\rho}{\kappa}\right ).
\end{equation}
Since the state of local neutrality $R^2=1$ is incompatible with the boundary condition $R=0$ at the axis of the vortex (\ref{Bernoulli_dimensionless}), the limiting solution given by Eqs.(\ref{Abrikosov}) and (\ref{limiting_potential}) only applies for sufficiently large $\rho$ to be determined below.  We stress that the existence of this solution only depends on two fundamental factors: the quantization of magnetic flux (\ref{flux_quantization}) and the interaction function $V(n)$ having a minimum at $n$ finite, $V'(n_0)=0$; the parabolic form (\ref{minimum}) is not a prerequisite.  Indeed substituting $n=n_{0}$ in Eq.(\ref{Bernoulli_modified}) we find $e\phi=-m\mathbf{v}^{2}/2$, which, in view of the expression for the flow velocity (\ref{vortex_velocity}) and the definitions of $w$ (\ref{conjecture}) and the coherence length (\ref{coherence_length}), recovers the expression for the limiting potential in the dimensionless form (\ref{limiting_potential}).  Therefore the limiting solution given by Eqs.(\ref{Abrikosov}) and (\ref{limiting_potential}) has a universal character.  

Another notable feature of the limiting solution (\ref{limiting_potential}) lies in the character of screening:  in the limit of zero Debye screening length $\gamma$, the electric field is screened over a length scale set by the London penetration depth $\kappa$:  $w(\rho\gg\kappa)=-(\pi l^{2}/2\kappa\rho)\exp(-2\rho/\kappa)$ \cite{NIST}.

\begin{figure}
\begin{center}
\includegraphics[width=\columnwidth]{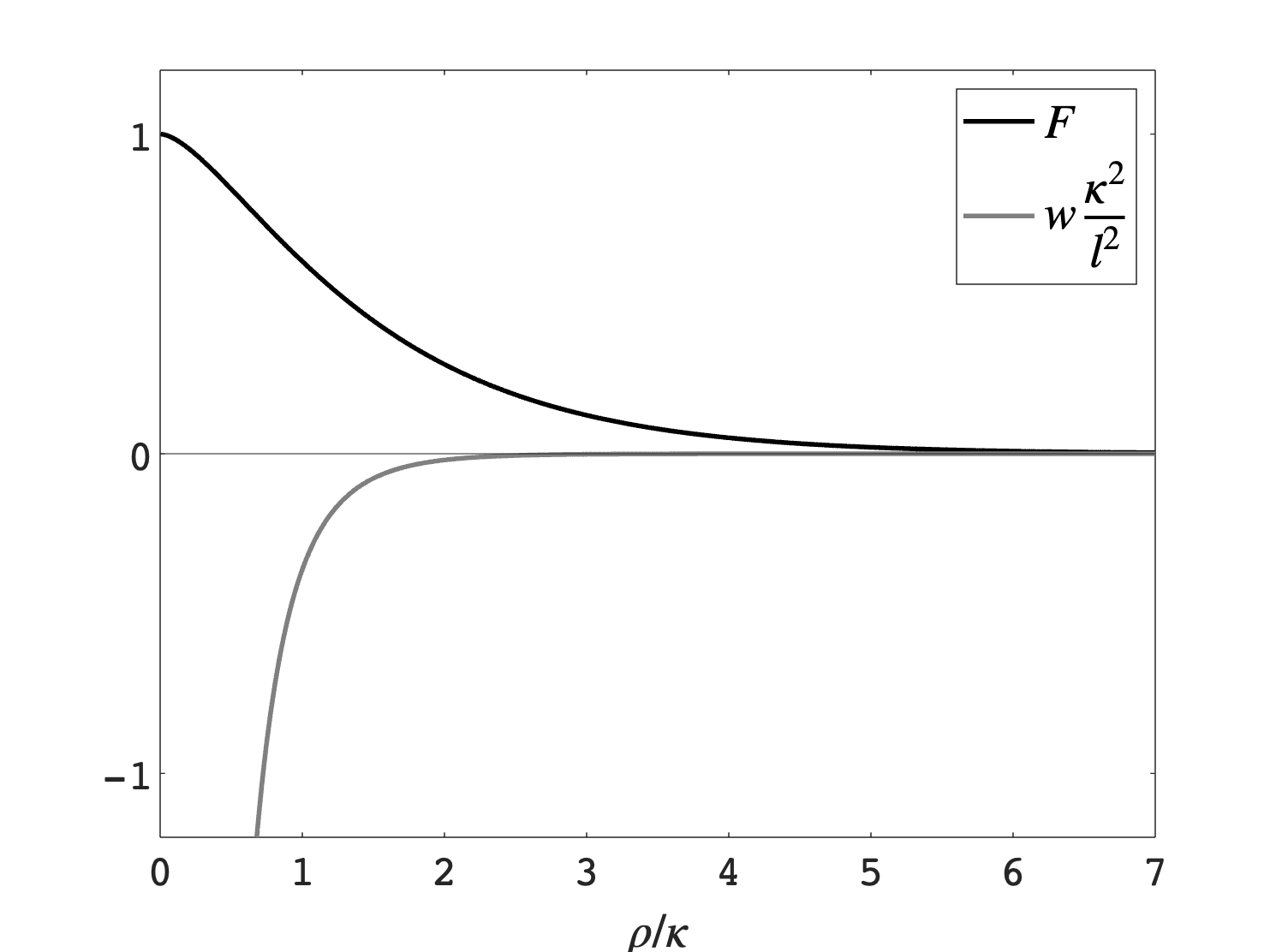}
\caption{The universal forms (in the strong-screening limit $\gamma\rightarrow 0$) of the magnetic field (\ref{magnetic_field}) (characterized by the function $F(\rho)$) and the electrostatic potential (\ref{conjecture}) (proportional to $w(\rho)$) associated with the Abrikosov flux line as given by Eqs.(\ref{Abrikosov}) and (\ref{limiting_potential}).  The length $\rho$ is measured in units of the coherence length (\ref{coherence_length}).}
\label{universal}
\end{center}
\end{figure}

Since formally the limiting solution (\ref{limiting_potential}) can be extended all the way to $\rho=0$, we must also resolve the puzzle of its very existence.  Indeed, local neutrality, $R^{2}=1$, means there are no uncompensated charges which implies that the electric field must be zero. We notice however that, according to the definitions of the potential $w$ (\ref{conjecture}) and the Debye screening length $\delta$ (\ref{London_Debye}), the physical scalar potential is proportional to $\delta^{2}$, i.e., $\phi\propto \delta^{2}w$. Since the $\gamma\rightarrow 0$ limit corresponds to the $\delta\rightarrow 0$ limit, the scalar potential $\phi$ vanishes as $\delta\rightarrow 0$ as expected when local neutrality is maintained. Consequently, the solution (\ref{limiting_potential}) should be understood as a statement that $\lim_{\delta\rightarrow 0}(\phi/\delta^{2})$ is finite.  The universal strong screening limiting solution given by Eqs.(\ref{Abrikosov}) and (\ref{limiting_potential}) is plotted in Figure \ref{universal}.

To match the limiting solution with the appropriate small-distance behavior, given by Eqs.(\ref{small_distance_density}), (\ref{small_distance_F}) and (\ref{small_distance-potential}), the constraint of perfect screening $\gamma\rightarrow 0$ must be lifted. Since $\kappa\gg 1$, we can restrict ourselves to distances $\rho\ll\kappa$ where $F=1$. In this case, Eq.(\ref{limiting_potential}) simplifies to 
\begin{equation}
\label{universal_potential}
w(\rho)=-\frac{l^{2}}{\rho^{2}}.
\end{equation}
Substituting this into the Poisson equation (\ref{Poisson_dimensionless}), we obtain an expression for the density $R^{2}$ that accounts for the deviation from the local neutrality limit to the second order in $\gamma$: 
 \begin{equation}
\label{correction2local_neutrality}
R^{2}(\rho)=1+\frac{4\gamma^{2}l^{2}}{\rho^{4}}.
\end{equation}
Comparing the first and second terms and ignoring the integer $l$, we deduce that the strong screening approximation and thus the limiting solution (\ref{Abrikosov}) and (\ref{limiting_potential}) apply up to distances of the order $\sqrt{\gamma}$. Consequently, the net (dimensionless) charge (per unit length) accumulated at distances $\rho\gtrsim\sqrt{\gamma}$ can be estimated as $-\gamma$. It is compensated by the net positive charge of the same order $\gamma$ for $\rho\lesssim\sqrt{\gamma}$ to ensure global neutrality. Therefore, the length scale $\sqrt{\gamma}$ has a meaning of the size of the vortex core. Since $\gamma\ll 1$, the size of the vortex core is substantially smaller than that in the weak screening regime. In physical units, the size of the vortex core in the strong screening regime is $\xi\sqrt{\gamma}=\xi\sqrt{\delta/\xi}=\sqrt{\delta\xi}$, whereas it is simply $\xi$ when electrostatic screening is weak (or electric effects are disregarded).      

Eq.(\ref{correction2local_neutrality}) also shows that as the distance from the vortex core increases, the superconductive density gradually approaches its bulk value from above. In conjunction with the small-distance behavior (\ref{small_distance_density}), this implies that the $R^{2}(\rho)$ dependence has a maximum. It appears plausible that this maximum occurs at $\rho\simeq \sqrt{\gamma}$, which, in light of Eq.(\ref{correction2local_neutrality}), suggests that the maximum density is independent of $\gamma$. 

Matching the small (\ref{small_distance_density}) and large (\ref{correction2local_neutrality}) distance dependences at $\rho\simeq\sqrt{\gamma}$ allows to write Eq.(\ref{small_distance_density}) more accurately as $R(\rho)\simeq (\rho/\sqrt{\gamma})^{|l|}$ valid for $\gamma\ll 1$.  Likewise, evaluating the potential (\ref{universal_potential}) at $\rho\simeq\sqrt{\gamma}$, the constant $w_{0}$ in Eq.(\ref{small_distance-potential}) can be estimated as $w_{0}\simeq -1/\gamma$.

While the expression for the function $F$ (\ref{Abrikosov}) (and ultimately the magnetic field (\ref{magnetic_field})) shares the same form with that in the Abrikosov theory \cite{Abrikosov}, their ranges of applicability differ significantly. In the strong screening limit $\gamma\rightarrow 0$, Eq.(\ref{Abrikosov}) holds true provided $\rho\gg \sqrt{\gamma}$. Conversely, with electric effects excluded, $w=0$, Eq.(\ref{Abrikosov}) holds if $\rho\gg 1$. This can be understood by noticing that the Abrikosov solution (\ref{Abrikosov}) has its origin in the observation that $R^{2}=1$ for $\rho\gg 1$.  However $R^{2}=1$ is also the condition of local neutrality that electric effects (absent in the Abrikosov theory) additionally enforce. Consequently, with electric effects included, Eq.(\ref{Abrikosov}) holds at smaller distances compared to those of the Abrikosov theory.  This reasoning applies regardless of the strength of screening.  Specifically, as the strength of screening increases from weak ($\gamma$ large) to strong ($\gamma$ small), the vortex core contracts.  Simultaneously, the maximum of the superconductive density increases in magnitude and shifts towards smaller $\rho$.  The same conclusion was already reached in our analysis of the regime of weak screening.

Contraction of the vortex core with increase of the strength of screening directly impacts the line energy of the vortex, Eq.(\ref{superconductor_energy_per_unit_length}). Similar to the Abrikosov theory \cite{LL9,Abrikosov}, for $\kappa\gg 1$ the dominant contribution to the line energy (\ref{superconductor_energy_per_unit_length}) arises from the kinetic energy of the superconductive electrons, which is represented by the $mn\mathbf{v}^{2}/2$ term within the integrand. Indeed, for $\kappa=\infty$, one finds that $F=1$ (\ref{Abrikosov}), the flow velocity (\ref{vortex_velocity}) falls off as $1/\rho$, and the line energy exhibits an infrared logarithmic divergence.    For $\kappa\gg 1$ and finite (and temporarily reverting to the physical units of length) we find
\begin{eqnarray}
\label{linear_energy}
\varepsilon_{l}&=&\int\frac{mn\mathbf{v}^{2}}{2}dxdy=\frac{\pi\hbar^{2}}{m}l^{2}\int_{0}^{\infty}\frac{n(\rho)F^{2}(\rho)}{\rho}
d\rho\nonumber\\
&\approx&\frac{\pi\hbar^{2}n_{0}}{m}l^{2}\int_{\sqrt{\delta\xi}}^{\lambda}\frac{d\rho}{\rho}=\frac{\pi\hbar^{2}n_{0}}{m}l^{2}\ln\frac{\kappa}{\sqrt{\gamma}}.
\end{eqnarray}
The lower integration limit in the third representation (tailored to the limit of strong screening $\gamma\ll 1$) arises from the density $n(\rho)$ decreasing from $n_{0}$ to zero within the vortex core of size $\sqrt{\delta\xi}$. Similarly, the upper integration limit reflects the fact that the function $F(\rho)$ (\ref{Abrikosov}) exponentially decreases for $\rho\gg\lambda$. The logarithmic nature of the divergence makes the precise choice of integration limits unimportant. The expression (\ref{linear_energy}) has logarithmic accuracy: in addition to the conditions $\kappa\gg 1$ and $\gamma\ll1$, one also requires that $\ln(\kappa/\sqrt{\gamma})\gg 1$.  Compared to the predictions of the Abrikosov theory in the $\ln\kappa\gg 1$ limit (or when electrostatic screening is weak), the line energy in the limit of strong screening is $\ln(\kappa/\sqrt{\gamma})/\ln\kappa$ times larger. This leads to a larger lower critical magnetic field, $H_{c_{1}}=4\pi\varepsilon_{1}/\Phi_{0}$, that marks the onset of penetration of the $l=\pm1$ vortices into type-II superconductor \cite{LL9,Abrikosov}. 

Contraction of the vortex core with increase of the strength of electrostatic screening also leads to larger (compared to Abrikosov's theory) magnetic field at small distances.  Indeed, evaluating Eq.(\ref{magnetic_field_small_rho}) at the edge of the vortex core $\rho\simeq \sqrt{\gamma}$ also provides an estimate for the magnetic field at the axis of the vortex,
\begin{equation}
\label{field_at_axis}
H(0)=\frac{\hbar cl}{e\lambda^{2}}\ln\frac{\kappa}{\sqrt{\gamma}}
\end{equation}
that enters Eq.(\ref{small_distance_F}).   As in the case of the line energy of the vortex (\ref{linear_energy}), this field is $\ln(\kappa/\sqrt{\gamma})/\ln\kappa$ times larger than its counterpart in the Abrikosov theory \cite{Abrikosov} (or if the electrostatic screening is weak).  

\subsection{Numerical results}

The problem of field configurations due the Abrikosov flux line accounting for electrostatic effects was also studied numerically.  Since $\kappa\gg 1$ and most significant deviations from Abrikosov's original predictions \cite{Abrikosov} take place at small distances from the vortex axis, $\rho\lesssim \kappa$, we carried out a detailed study of the purely electric version of the problem, $\kappa=\infty$.  Then, according to Amp\`ere's law (\ref{Ampere_dimensionless}), we only need to solve the $F=1$ case of Eq.(\ref{Bernoulli_dimensionless}) together with the Poisson equation (\ref{Poisson_dimensionless}); a code is available online \cite{Mia}.  
\begin{figure*}
\begin{center}
\includegraphics[width=2\columnwidth]{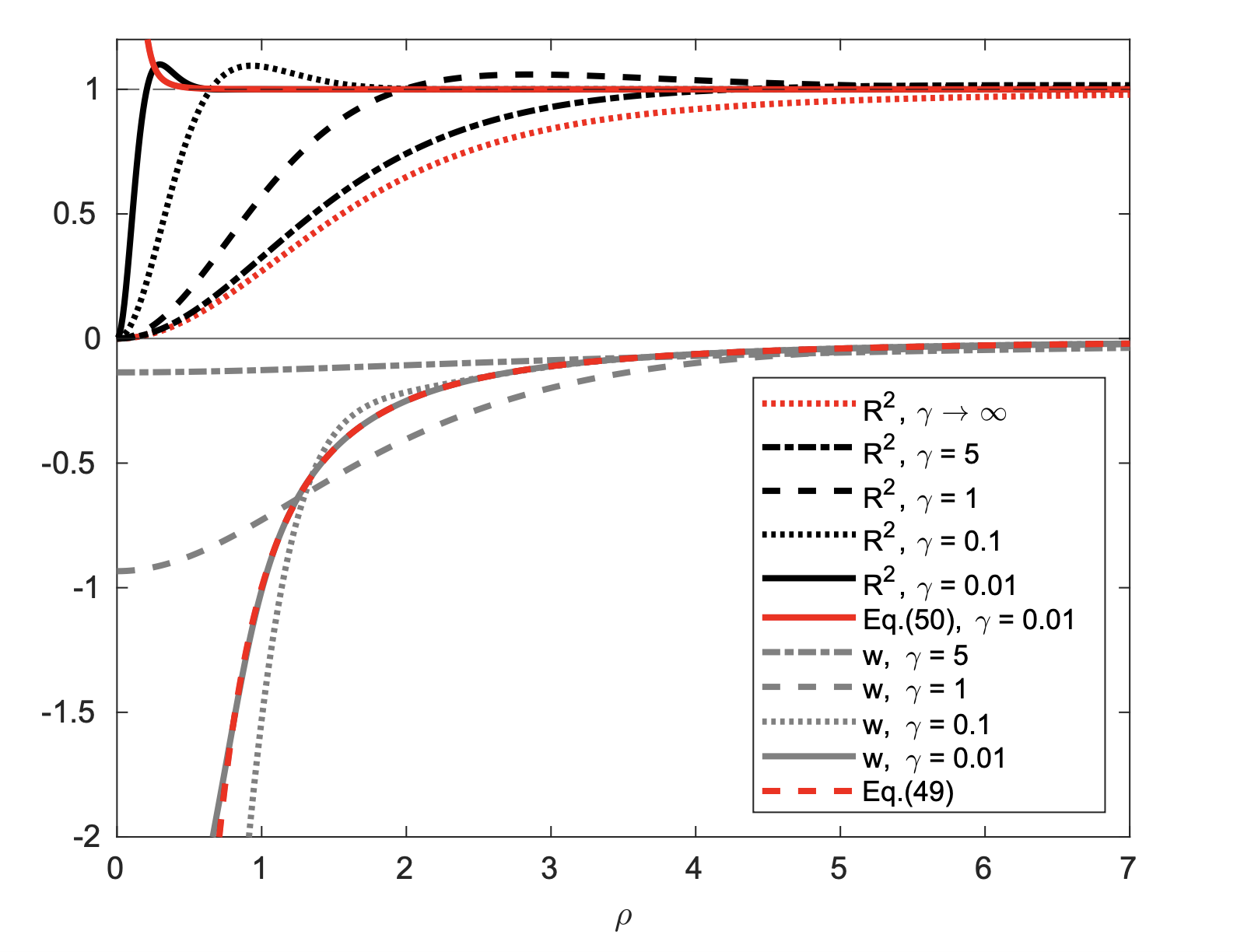}
\caption{Dimensionless density of superconductive electrons $R^{2}(\rho)$ (bold) and potential $w(\rho)$ (grey scale) of an $|l|=1$ Abrikosov vortex at distance $\rho\ll\kappa$ from its axis according to Eqs.(\ref{Bernoulli_dimensionless}),(\ref{Poisson_dimensionless}) and (\ref{Ampere_dimensionless}) in the $\kappa\gg 1$ limit for different strengths of electrostatic screening parameterized by the dimensionless Debye screening length $\gamma$.  Dotted red curve is the Ginzburg-Pitaevskii solution for the density profile $R^{2}(\rho)$ of the Onsager-Feynman vortex in neutral superfluid \cite{LL9}.  Solid red and dashed red curves represent asymptotic strong screening results for the density and potential given by Eqs.(\ref{correction2local_neutrality}) and (\ref{universal_potential}), respectively. }
\label{numerical}
\end{center}
\end{figure*}

In the limits of weak $\gamma\gg 1$ and strong $\gamma\ll 1$ electrostatic screening our results confirm the physical reasoning and specific predictions of the asymptotic analysis.  Additionally they supply valuable information about the cases of moderate screening $\gamma\simeq 1$, intermediate distances and magnitude of the effect of enhanced density of superconductive electrons outside the vortex core.  Below we limit ourselves to the practically relevant case of $|l|=1$ corresponding to the Abrikosov vortex carrying single flux quantum.

The results are assembled in Figure \ref{numerical}, and it is useful to start their discussion  with monotonically increasing red dotted curve.  It corresponds to the Ginzburg-Pitaevskii solution for the density profile $R^{2}(\rho)$ of the Onsager-Feynman vortex in a neutral superfluid \cite{LL9}.  It is a solution to the $F=1$, $w=0$ version of Eq.(\ref{Bernoulli_dimensionless}) and simultaneously the limit of infinitesimally weak screening, $\gamma\rightarrow \infty$, when electrostatic effects are absent and the potential is zero.  As soon as $\gamma$ becomes finite still remaining in the weak screening regime $\gamma\gg 1$, the density profile $R^{2}(\rho)$ develops a maximum at $\rho$ large above the bulk value $R^{2}=1$;  the latter (for $\rho$ large) is approached from above.  Additionally, there is a steepening of the density profile at small distances that points to shrinking of the vortex core;  electrostatic effects in the form of finite potential $w(\rho)$ are now clearly present.  

As the strength of screening increases, i.e. $\gamma$ gets smaller, the maximum of $R^{2}(\rho)$ becomes larger and displaces closer to the vortex axis.  The density profile steepens at small distances even more, and the potential $w(\rho)$ deepens still.  At an intermediate screening strength of $\gamma=1$ the maximum superconductive density is about $5\%$ higher than the bulk value.

These tendencies persist as one moves into the regime of strong screening $\gamma\le 1$.  What becomes particularly clear in this regime is a trend of the density of superconductive electrons to adhere to its bulk value at ever smaller distances.  As the strength of screening increases, i.e. $\gamma$ gets smaller, the maximum density increases too.  At $\gamma=0.01$ the maximum superconductive density is about $10\%$ higher than the bulk value.  Additionally, as the strength of screening increases, the density profile $R^{2}(\rho)$ and the potential $w(\rho)$ approach their asymptotic limits given by Eqs.(\ref{correction2local_neutrality}) (red solid curve) and (\ref{universal_potential}) (red dashed curve), respectively, everywhere excepting a narrow vicinity of the vortex axis constituting the vortex core.        

\section{Conclusions}

To summarize, we demonstrated that at zero temperature electrostatic effects in tightly bound Cooper pair superconductors qualitatively modify the structure of the Abrikosov flux line.  The latter is now characterized by profiles of the magnetic and electric fields and the density of superconductive electrons.  The latter exhibits a maximum that exceeds the bulk value of the superconductive density.  While superconductivity is still degraded within the (smaller) vortex core, it is enhanced outside of it.  The underlying physical mechanism behind this behavior is electrostatic screening that enforces global neutrality.  We also demonstrated that in the limit of strong electrostatic screening profiles of the magnetic and electric fields attain a universal form that has its origin in quantization of the magnetic flux and existence of equilibrium bulk density of superconductive electrons.

We fully expect that our conclusions will remain largely intact under less strict conditions (non-zero but low temperature and moderately bound Cooper pairs) when there is some small presence of normal electrons.  

We are aware of only two examples of charged topological defects.  One of them is dislocations in ionic crystals \cite{Esh}.  They acquire charge due to difference in energy required to form positive and negative ion vacancies.  This line charge is screened by a Debye-H\"uckel cloud of oppositely charged vacancies vaguely resembling the problem studied in our work.

Our second example is that of dyons - hypothetical particles carrying both magnetic and electric charges \cite{Schwinger}.  Since quantized magnetic flux of the Abrikosov vortex is essentially the magnetic charge, there certainly is a similarity with the problem studied in our work.  One of the central differences however is that magnetic and electric fields in a dyon are unscreened.   

We hope experimental efforts will be undertaken to determine the detailed structure of individual Abrikosov flux lines and compare with our predictions.  The most promising experimental technique appears to be the scanning tunneling microscopy (STM) \cite{STM1,STM2}.  What is measured is the local density of states (LDOS) at different energies, which reflects the suppression of the superconducting gap (i.e. the density of superconductive electrons) as the vortex axis is approached.  STM's atomic scale spatial resolution makes is optimally suited to carry out detailed measurements of the profile of the density of superconductive electrons.  Having determined the superconductive density, the profiles of the electric and magnetic fields can be inferred from the Poisson equation (\ref{Poisson_dimensionless}) and Amp\`ere law (\ref{Ampere_dimensionless}), respectively.  

Suitable materials include both conventional superconductors like \textrm{NbSe}$_{2}$, \textrm{Pb}, and copper-based superconductors belonging to the Bismuth Strontium Calcium Copper oxide family. For instance, since \textrm{NbSe}$_{2}$ is a good conductor, electrostatic screening is exceptionally strong and occurs over a very short distance, approximately $\delta \simeq 0.1~nm$. Analysis and images of the original observations \cite{STM1} are consistent with a coherence length $\xi \simeq 10~nm$. Consequently, the dimensionless Debye screening length $\gamma = \delta/\xi$ (\ref{Poisson_dimensionless}) is of the order $0.01$. Our theory then predicts that the density of superconductive electrons outside the vortex core is about $10\%$ higher than its bulk counterpart.

Existence of the electric field profile that accompanies the Abrikosov flux line opens up a possibility to control vortex motion with an electric field.  Incidentally, an ability to control movement of charged dislocations by an electric field was demonstrated recently \cite{Dance}.  

\section{Acknowledgements}

We are grateful to J. P. Straley and G. E. Volovik for valuable suggestions and constructive criticism.

\end{document}